\documentstyle[12pt]{article}

\newcommand{\nc}{\newcommand}
\newcommand{\noi}{\noindent}
\def\sq{\hbox {\rlap{$\sqcap$}$\sqcup$}}
\newtheorem{lemma}{Lemma}[section]
\newtheorem{theorem}[lemma]{Theorem}

\newtheorem{remark}[lemma]{Remark}
\newtheorem{example}[lemma]{Example}

\newtheorem{definition}[lemma]{Definition}

\nc{\QED}{\mbox{}\hfill \raisebox{-2pt}{\rule{5.6pt}{8pt}\rule{4pt}{0pt}}
          \medskip\par}

\nc{\N}{{\rm I\mkern-4.0mu N}}
\nc{\R}{{\rm I\mkern-4.0mu R}}
\nc{\Z}{{\sf Z\mkern-6.5mu Z}}
\nc{\Id}{I}

\nc{\di}{\displaystyle}
\nc{\beq}{\begin{equation}}
\nc{\edq}{\end{equation}}
\nc{\beqn}{\begin{eqnarray}}
\nc{\edqn}{\end{eqnarray}}

\nc{\C}{{\mathchoice {\setbox0=\hbox{$\displaystyle\rm C$}\hbox{\hbox
to0pt{\kern0.4\wd0\vrule height0.9\ht0\hss}\box0}}
{\setbox0=\hbox{$\textstyle\rm C$}\hbox{\hbox
to0pt{\kern0.4\wd0\vrule height0.9\ht0\hss}\box0}}
{\setbox0=\hbox{$\scriptstyle\rm C$}\hbox{\hbox
to0pt{\kern0.4\wd0\vrule height0.9\ht0\hss}\box0}}
{\setbox0=\hbox{$\scriptscriptstyle\rm C$}\hbox{\hbox
to0pt{\kern0.4\wd0\vrule height0.9\ht0\hss}\box0}}}}

\nc{\Sph}{{\mathchoice
{\setbox0=\hbox{$\displaystyle     \rm S$}\hbox{\raise0.5\ht0\hbox
to0pt{\kern0.35\wd0\vrule height0.45\ht0\hss}\hbox
to0pt{\kern0.55\wd0\vrule height0.5\ht0\hss}\box0}}
{\setbox0=\hbox{$\textstyle        \rm S$}\hbox{\raise0.5\ht0\hbox
to0pt{\kern0.35\wd0\vrule height0.45\ht0\hss}\hbox
to0pt{\kern0.55\wd0\vrule height0.5\ht0\hss}\box0}}
{\setbox0=\hbox{$\scriptstyle      \rm S$}\hbox{\raise0.5\ht0\hbox
to0pt{\kern0.35\wd0\vrule height0.45\ht0\hss}\raise0.05\ht0\hbox
to0pt{\kern0.5\wd0\vrule height0.45\ht0\hss}\box0}}
{\setbox0=\hbox{$\scriptscriptstyle\rm S$}\hbox{\raise0.5\ht0\hbox
to0pt{\kern0.4\wd0\vrule height0.45\ht0\hss}\raise0.05\ht0\hbox
to0pt{\kern0.55\wd0\vrule height0.45\ht0\hss}\box0}}}}

\begin{document}

\begin{center}
{\bf \Large A Proof of the Gutzwiller Semiclassical Trace Formula}\\
[3 truemm]
{\bf \Large  Using Coherent
States Decomposition} \\
[8 truemm] 
{\bf Monique Combescure}\\
Laboratoire de Physique Th\'eorique et Hautes
Energies\footnote{Laboratoire associ\'e au Centre National de la Recherche
Scientifique - URA D0063} \\  Universit\'e de Paris XI, B\^atiment 210, F-91405
Orsay Cedex, France\\
[3 truemm]
{\bf James Ralston}\\
Department of Mathematics, UCLA, 
Los Angeles CA 90024-1555, U.S.A.\\
[3 truemm]
{\bf Didier Robert}\\
D\'epartement de Math\'ematiques, Universit\'e de Nantes\\
2, rue de la Houssini\`ere, F-44072 Nantes Cedex 03, France - URA CNRS 758
\end{center}

\abstract{
The  Gutzwiller trace formula links the eigenvalues of the
Schr\"odinger operator
$\widehat{H}$ as Planck's constant goes to zero (the semiclassical r\'egime)
with the closed orbits of the corresponding  classical mechanical  system.
Gutzwiller  gave a
heuristic proof of this trace formula,
using the Feynman integral representation for the propagator of
$\widehat{H}$. Later, using the theory of Fourier integral operators,
mathematicians gave rigorous proofs of the formula in various
settings.
Here we show how the use of coherent states
allows us to give a simple and direct proof.}

\section { Introduction}

Our goal in this paper is to give a simple proof of the
``semiclassical Gutzwiller trace formula''.
The pioneering works in quantum physics of Gutzwiller \cite{gu} (1971)
and Balian-Bloch \cite{babl1} \cite{babl2}
(1972-74) showed that the trace of a quantum observable $\widehat{A}$,
localized in a
spectral neighborhood of size $O(\hbar)$ of an energy $E$ for the quantum
Hamiltonian
$\widehat{H}$, can be expressed in terms of averages of the classical
observable $A$ associated with $\widehat{A}$ over invariant sets for the
flow of the classical Hamiltonian $H$ associated with $\widehat H$.
This is related to the spectral asymptotics
for $\widehat H$ in the semi-classical limit,
and it can be understood as a ``correspondence
principle''
between classical and quantum mechanics as Planck's constant $\hbar$ goes to
zero.

Between 1973 and 1975 several authors gave rigorous derivations of related
results, generalizing the classical Poisson summation formula
from $d^2/ d\theta^2$ on the circle to elliptic operators
on compact manifolds:
Colin de Verdi\`ere \cite{cdv},
Chazarain \cite{ch}, Duistermaat-Guillemin
\cite{dugu}. The first article is based on
a parametrix construction for the associated heat equation, while the
second two replace this with a parametrix, constructed as a Fourier
integral operator,
for the associated wave
equation.
More recently, papers by Guillemin-Uribe (1989), Paul-Uribe (1991, 1995),
Meinrenken
(1992) and Dozias (1994) have developed the necessary tools from microlocal
analysis
in a nonhomogeneous (semiclassical)
setting to deal with
Schr\"odinger-type Hamiltonians. Extensions and simplifications
of these methods have been given by Petkov-Popov \cite{pepo},
and Charbonnel-Popov \cite{chpo}.

The coherent states approach presented here seems particularly suitable
when one wishes to compare the phase space quantum
picture  with the phase space classical flow. Furthermore, it
avoids problems with caustics, and the Maslov indices appear naturally.
In short, it implies the Gutzwiller
trace formula in a very simple  and transparent way,
without any use of the global theory of Fourier integral operators. In
their place we use
the coherent states approximation (gaussian beams)
and the stationary phase theorem .

The use of gaussian wave packets is such a useful idea that one
can trace it back to the very beginning of quantum mechanics, for
instance,
Schr\"odinger \cite{sch} (1926). However, the realization that these
approximations are universally applicable, and that they are valid
for arbitrarily long times, has developed gradually.
In the mathematical literature these approximations have
never become textbook material, and this has lead to their repeated
rediscovery with a variety of different names, e.g. coherent states and
gaussian beams.
The first place that we have found where they are
used in some generality is Babich \cite{ba} (1968)
(see also \cite{babu}). Since then
they have appeared, often as independent discoveries, in the work of
Arnaud \cite{ar} (1973), Keller \cite{ke}
(1974),
Heller \cite{he} (1975, 1987), Ralston \cite{ra1},\cite{ra2} (1976,1982),
Hagedorn
\cite{ha} (1980-85), and Littlejohn \cite{li} (1986) -- and probably
many more that we have not found.
Their use in
trace formulas was proposed by Wilkinson \cite{wi} (1987).
The propagation formulas of \cite{ha} were extended
in Combescure-Robert \cite{coro2}, with
a detailed estimate on the error both in time and in Planck's constant.
The early application of these methods in \cite{ba} was for the
construction of quasi-modes, and this has been pursued further in
\cite{ra1} and Paul-Uribe \cite{paur1}. There have also been recent
applications to the pointwise behaviour of semiclassical measures
\cite{paur2}.

{\bf Acknowledgement}:  The authors thank J. Sj\"ostrand for helpful
discussions of this topic, and J. Ramanathan for valuable comments on
the preliminary version of this paper.

\section{ The Semiclassical Gutzwiller Trace Formula }

We consider a quantum system in $L^2(\R^n)$ with Hamiltonian

\beq
\widehat{H} = - \hbar^2 \Delta + V(x),
\edq
 where $\Delta$ is the Laplacian in $L^2(\R^n)$ and $V(x)$ a real,
$ C^{\infty}(\R^n)$ potential.
 The corresponding Hamiltonian for the classical motion is

$$H(q, p) = p^2 + V(q), $$
\noindent and for a given energy $E$ ($\in \R$) we denote by $\Sigma_E$ the
``energy
shell''
\beq
\Sigma_E := \left\{ (q, p) \in \R^{2n} \, : \, H(q, p) = E \right\}.
\edq
More generally we shall consider Hamiltonians $\widehat{H}$ obained by the
$\hbar$-Weyl
 quantization of the classical Hamiltonian $H$, so that
 $\widehat{H} = Op^w_\hbar(H)$, where
\beq
  Op^w_\hbar(H)\psi(x) = ({2 \pi \hbar})^{-n} \int_{\R^{2n}}
 H\left( \frac{x + y}{2}, \xi \right)\psi(y){\rm e}^{\frac{i(x -
y)\cdot\xi}{\hbar}}dyd\xi
\edq
The Hamiltonian $H$ is assumed to be a smooth, real valued function of
$ z = (x, \xi)\in\R^{2n}$,
and to satisfy the following  global estimates
\begin{itemize}
\item{(H.0)} there exist non-negative constants $C, m, C_\gamma$
  such that
\beqn
\vert \partial^\gamma_z H(z)\vert \leq C_\gamma <H(z)>,\;\;\forall
z\in\R^{2n}, \;
\forall \gamma\in  \N^{2n}\\
<H(z)> \;\;\leq\;\; C<H(z^\prime)>\cdot <z-z^\prime>^m,\;\; \forall z,\;
z^\prime \in
\R^{2n}
\edqn
where we have used the  notation $<u> = (1+\vert u\vert^2)^{1/2}$ for
$u\in \R^m$.
\end{itemize}
\begin{remark} i) $H(q, p) = p^2 + V(q) $ satisfies {\rm (H.0)},
 if $V(q)$  is bounded below by some $a>0$ and satisfies  the property
 {\rm (H.0)} in the variable $q$.\\
ii) The technical condition {\rm (H.0)} implies in particular  that
$\widehat{H}$  is
 essentially self-adjoint
on $L^2(\R^n)$ for $\hbar$ small enough and  that
 $\chi(\widehat{H})$ is a $\hbar$-pseudodifferential operator
 if $\chi \in C_0^{\infty}(\R)$ (see \cite{hero}) .
\end{remark}

 Let us denote by $\phi_t$ the classical flow induced by Hamilton's
equations with Hamiltonian $H$ , and by $S(q,p;t)$ the classical action
along the
trajectory starting at $(q, p)$ at time $t = 0$, and evolving during time
$t$:

\beq
S(q,p;t) = \int_0^t  \left ( p_s \cdot \dot{q}_s - H(q, p) \right )ds
\edq
 where $(q_t, p_t) = \phi_{t}(q, p)$, and dot denotes the
derivative with respect to time.  We shall also use  the notation:
$\alpha_t = \phi_t(\alpha)$ where  $ \alpha = (q, p) \in \R^{2n}$,
is a phase space point.

An important
role in
what follows is played by the ``linearized flow'' around the classical
trajectory,
which is defined as follows. Let
\beq
H^{\prime\prime}(\alpha_t) = \left . {\partial^2 H \over \partial \alpha^2}
\right |_{\alpha =
\alpha_t}
\edq
 be the Hessian of $H$ at point $\alpha_t = \phi_t(\alpha )$ of the
classical trajectory.  Let $J$
be the
symplectic matrix

\beq
J = \pmatrix{0 &\quad I \cr
-I &\quad 0 \cr}
\edq
 where 0 and $I$ are respectively the null and identity $n \times n$
matrices. Let $F(t)$ be the $2n \times 2n$ real symplectic matrix
solution of
the linear  differential equation

\beq\label{met}
\cases{
\dot{F}(t) = J \ H^{\prime\prime}(\alpha_t) \ F(t) \cr
\cr
F(0) = \pmatrix{I &0 \cr 0 &I \cr
} = \Id \cr}
\edq
 $F(t)$ depends on $\alpha = (q,p)$, the initial
point for the classical trajectory, $\alpha_t$.

Let $\gamma$ be a \underbar{closed} orbit on $\sum_E$ with period
$T_{\gamma}$, and let us denote simply by $F_{\gamma}$ the matrix
 $F_{\gamma} = F(T_{\gamma}) $.
\noindent $F_{\gamma}$ is usually called the ``monodromy matrix'' of the
closed
orbit $\gamma$. Of course, $F_{\gamma}$ does depend on $\alpha$, but its
eigenvalues do not, since the monodromy matrix with a different initial
point on $\gamma$ is conjugate
to $F_{\gamma}$.
$F_{\gamma}$ has 1 as eigenvalue of algebraic
multiplicity at least equal to 2. In all that follows, we shall use the
following definition
 \begin{definition}
  We say that $\gamma$ is  a nondegenerate orbit if the eigenvalue 1 of
$F_{\gamma}$ has algebraic multiplicity 2.
\end{definition}
Let $\sigma$ denote the usual symplectic form on $\R^{2n}$

\beq
\sigma (\alpha , \alpha ') = p \cdot q' - p' \cdot q  \qquad \matrix{\alpha =
(q, p);\; \alpha ' = (q', p') \cr}
\edq
 ($\cdot$ is usual scalar product in $\R^n$). We denote by $\{ \alpha_1,
\alpha '_1\}$ the eigenspace of $F_{\gamma}$ belonging to the eigenvalue 1,
and by
${\it V}$ its orthogonal complement in the sense of the symplectic form
$\sigma$
\beq
{\it V} = \left \{ \alpha \in \R^{2n} \ : \ \sigma (\alpha , \alpha_1)
=\sigma (\alpha , \alpha '_1 ) = 0 \right \} \ \ \ .
\edq

\noindent Then, the restriction $P_{\gamma}$ of $F_{\gamma}$ to ${\it V}$ is
called the  (linearized) ``Poincar\'e map'' for $\gamma$.

In some cases the Hamiltonian flow will contain manifolds of periodic
orbits with the same energy. When this happens, the periodic orbits
will necessarily be degenerate, but the techniques we use here can
still apply. The precise hypothesis for this (\lq\lq Hypothesis C")
will be given in Section 4. Following Duistermaat and Guillemin we call
this a \lq\lq clean intersection hypothesis", but it is more explicit
than other versions of this assumption. Since the statement of the
trace formula is simpler and more informative when one does assume
that the periodic orbits are nondegenerate, we will give that formula
here.

We shall now assume the following. Let $(\Gamma_E)_T$ be the set of all
periodic
orbits on $\sum_E$ with periods $T_\gamma$,  $ 0 < \vert T_{\gamma}\vert
\leq T$ (including repetitions of
primitive orbits and assigning negative periods to primitive orbits
traced in the opposite sense). Then we require:
\begin{itemize}
\item{(H.1)} There exists $\delta E > 0$ such that
\hbox{$ H^{-1}([E-\delta E, E+\delta E])$} is  a compact set of $\R^{2n}$
and
$E$ is a noncritical value of $H$ (i.e. $H(z) = E \Rightarrow \nabla
H(z) \not= 0$).
\item{(H.2)} For any $T > 0$, $(\Gamma_E)_T$ is a discrete set, with periods
$ -T \leq T_{\gamma_1} < \cdots < T_{\gamma_N} \leq T$.
\item{(H.3)} All $\gamma$ in $(\Gamma_E)_T$ are nondegenerate, i.e. 1
is not an eigenvalue for the
corresponding ``Poincar\'e map'', $P_{\gamma}$.\\
We can  now  state the  Gutzwiller trace formula. Let
$\widehat{A} = Op^w_\hbar(A)$ be a quantum observable, such that $A$
satisfies  the following
\item{(H.4)}
 there exists  $\delta \in \R$,  $C_\gamma>0$ ($\gamma \in\N^{2n}$),
  such that
$$
\vert \partial^\gamma_z A(z)\vert \leq C_\gamma <H(z)>^\delta\;\;
\forall z\in  \R^{2n},
$$
\item{(H.5)}   $g$ a ${\cal C}^{\infty}$ function whose
Fourier transform $\widehat{g}$ is of compact support with \\
\qquad ${\rm Supp} \ \widehat{g} \subset [-T, T]$
\item{} and let $\chi$ be a smooth function with a compact support
contained in
$] E - \delta E , E + \delta E [$, equal to 1 in a neighborhood of $E$.
Then the
following ``regularized density of states'' $\rho_A(E)$ is
well
defined
\end{itemize}
\beq
\rho_A(E) = {\rm Tr} \left( \chi (\widehat{H}) \widehat{A} \chi
(\widehat{H}) g
\left( {E - \widehat{H} \over \hbar} \right)\right)
\edq
Note that (H.1) implies that the spectrum of $\widehat{H}$ is
purely discrete
  in a neighborhood of $E$ so that  $\rho_A(E)$ is well defined.
 Then we have the
following,

\begin{theorem}: Assume {\rm (H.0)-(H.3)} are satified for $H$, {\rm (H.4)}
for $A$ and {\rm (H.5)} for $g$. Then the following
 asymptotic expansion  holds true, modulo $O(\hbar^\infty)$,
\beqn\label{GTF}
\rho_A(E) \equiv
(\pi)^{-n/2}\widehat{g}(0)\hbar^{-(n-1)} \int_{\Sigma_E} A(\alpha )
d\sigma_E (\alpha )  +
\sum_{k \geq - n + 2} c_k (\widehat{g}) \hbar^k   \\ \nonumber
+ \sum_{\gamma \in (\Gamma_E)_T}   (2\pi)^{n/2-1}\left\{
\widehat{g}(T_{\gamma})
{{{\rm e}^{i(S_{\gamma}/\hbar + \sigma_{\gamma} \pi /2)}}\over{|\det (\Id -
P_{\gamma})|^{1/2}}}
\int_0^{T^*_{\gamma}}A(\alpha_s)ds + \sum_{j \geq 1}
d_j^{\gamma}(\widehat{g}) \hbar^j \right\}
\edqn
 where $A(\alpha )$ is the classical Weyl symbol of $\widehat{A}$, \\
$T_{\gamma}^*$ is the primitive period of $\gamma$, \\
$\sigma_{\gamma}$ is the Maslov index of $\gamma$  ( $\sigma_{\gamma}\in
\Z$ ),\\
$S_{\gamma} = \oint_{\gamma} p dq$ is the classical action along
$\gamma$, \\
$c_k(\widehat{g})$   are distributions in $\widehat{g}$ with support in  $\{
0\}$, \\
$d_j^{\gamma}(\widehat{g})$ are distributions in $\widehat{g}$ with support
 $\{T_{\gamma}\}$
 and $d \sigma_E$ is the Liouville measure on $\sum_E$:

$$d\sigma_E = {d\Sigma_E \over |\nabla H|}
 \qquad (d\Sigma_E \ \hbox{ is the Euclidean
measure on}\ \Sigma_E)$$
\end{theorem}
\begin{remark}
We can include more general Hamiltonians depending explicitly in $\hbar$, \\
 $\displaystyle{H = \sum_{j=1}^K \hbar^jH^{(j)}}$ such that
 $H^{(0)}$ satisfies {\rm (H.0)} and for $j\geq 1$,
 \beq\label{sum}
 \vert\partial^\gamma H^{(j)}(z)\vert  \leq C_{\gamma,j}<H^{(0)}(z)>
\edq
 It is useful for applications to  consider Hamiltonians
 like $H^{(0)} + \hbar H^{(1)}$ where $H^{(1)}$  may
 be, for example, a spin term.  In that case the formula {\rm (\ref{GTF})}
is true
 with different coefficients. In particular the first term in the
contribution of
 $T_\gamma$ is multiplied by
 \hbox{$\exp\left(-i\int_0^{T^*_\gamma}H^{(1)}(\alpha_s)ds\right)$}.
\end{remark}
\begin{remark}
For Schr\"odinger operators we only need smoothness of the
 potential $V$. In this case the trace formula (\ref{GTF}) is still  valid
  without any   assumptions at infinity  for $V$ when  we
restrict ourselves
  to a compact energy surface, assuming
 \hbox{$ E < \liminf_{\vert x \vert \rightarrow \infty}V(x)$}.
Using  exponential decrease
 of the eigenfunctions \cite{hesj} we can prove that, modulo an error term
  of order $O(\hbar^{+\infty})$,  the potential $V$
 can be replaced by a potential $\tilde{V}$ satisfying the assumptions of
the Remark (2.1).
\end{remark}
\section{Preparations for the Proof}

We shall make use of ``coherent states'' which can be defined as
follows.  Let
\beq\label{bs}
\psi_0(x) = (\hbar \pi )^{-n/4} \ \exp \left(-\frac{ \vert x\vert
^2}{2\hbar}\right),
\edq
 be the ground state of the $n$-dimensional harmonic oscillator, and for
$\alpha = (q, p) \in \R^{2n}$,

\beq
{\cal T}(\alpha ) = \exp \left \{ {i \over \hbar} (p \cdot x - q \cdot
\hbar D_x) \right \}
\edq
 is the Weyl-Heisenberg operator of translation by $\alpha$ in phase space
where $D_x = \frac{\partial}{i\partial x}$
  We also denote by
\beq
\varphi_{\alpha} = {\cal T}(\alpha ) \psi_0
\edq
 the usual coherent states centered at the point $\alpha$.
Then it is known that any operator $B$ with a symbol decreasing sufficiently
rapidly is in trace class (see \cite{fo}),
and its trace can be computed by
\beq
{\rm Tr} B = (2\pi\hbar)^{-n} \int  <\varphi_{\alpha} , B
\varphi_{\alpha}>d\alpha.
\edq
The regularized density of states $\rho_A(E)$ can now be rewritten as
\beq
\rho_A(E) = (2\pi)^{-n-1}\hbar^{-n} \int  \widehat{g}(t) \ e^{iEt/\hbar} \
<\varphi_{\alpha}
, \widehat{A}_{\chi} \ U(t) \ \varphi_{\alpha}>dtd\alpha
\edq
 where $U(t)$ is the quantum unitary group~:
\beq
U(t) = e^{-it\widehat{H}/\hbar}
\edq
 and $\widehat{A}_{\chi} = \chi (\widehat{H}) \widehat{A} {\chi}
(\widehat{H})$.

Our strategy for computing the behavior of $\rho_A(E)$  as $\hbar$ goes to
zero
is first to compute
the bracket
\beq\label{bra}
m(\alpha, t) = <\widehat{A_\chi}\varphi_\alpha, U(t)\varphi_\alpha >,
\edq
where we drop the subscript $\chi$ in $A_\chi$ for simplicity.
It is useful to rewrite (\ref{bs}) as
\beq
\psi_0 = \Lambda_{\hbar} \widetilde{\psi}_0,
\edq
 where $\Lambda_{\hbar}$ is the following scaling operator:
\beq
\left ( \Lambda_{\hbar} \psi \right ) (x) = \hbar^{-n/4} \ \psi \left ( x
\hbar^{-1/2} \right ) \;\;
{\rm  and}\;\;
\widetilde{\psi}_0 (x) = \pi^{-n/4} \exp \left ( - \vert x\vert^2/2 \right
) \ \
\ .
\edq

 First of all we shall use the following lemma,  giving the action
  of an $\hbar$-pseudodifferential operator on a Gaussian.
\begin{lemma}
Assume  that $A$ satisfies {\rm (H.0)}. Then we have
\beq
\widehat{A} \varphi_{\alpha} =
\sum\limits_{\gamma} \hbar^{\frac{|\gamma|}{2}}\frac{ \partial^{\gamma}
A(\alpha)}{\gamma!}
\Psi_{\gamma , \alpha} + O(\hbar^{\infty})
\edq
in $L^2(\R^n)$, where  $\gamma  \in \N^{2n}$, $| \gamma | =
\sum\limits_{1}^{2n}$,
$\gamma ! = \prod\limits_{1}^{2n} \gamma_j !$ and
\beq
\Psi_{\gamma , \alpha} =
{\cal T}(\alpha ) \Lambda_{\hbar} Op^w_1(z^\gamma)\widetilde{\psi}_0.
\edq
 where  $Op^w_1(z^\gamma)$ is the 1-Weyl quantization of the monomial~:\\
$(x, \xi)^{\gamma} = x^{\gamma^\prime}\xi^{\gamma^{\prime\prime}}$,
\hbox{$\gamma = (\gamma^\prime, \gamma^{\prime\prime})\in\N^{2n}$}.
\end{lemma}
This lemma is easily proved using  a
scaling argument and Taylor expansion
for the symbol $A$
  around  the point $\alpha$.
Thus $m(t,\alpha)$ is a linear combination of terms like
\beq
m_\gamma( \alpha, t ) = <\Psi_{\gamma , \alpha}, U(t) \varphi_{\alpha}> .
\edq
 Now we compute $ U(t) \varphi_{\alpha}$,  using
 the semiclassical propagation of coherent states result
 as it was formulated in  Combescure-Robert \cite{coro2}.
 We recall that $F(t)$ is a time dependent symplectic matrix (Jacobi
matrix)
  defined by the linear equation (\ref{met}).
Met$F$  denotes the metaplectic representation of the
linearized flow $F$ (see for example Folland \cite{fo}), and
  the $\hbar$-dependent metaplectic representation is defined by
 \beq
{\rm Met}_\hbar(F) = \Lambda_\hbar^{-1}{\rm Met}(F)\Lambda_\hbar
\edq
  We will also use the notation
\beq
\delta(\alpha, t) = \int_0^t p_s\cdot\dot{q_s}ds -tH(\alpha)
-\frac{p_t\cdot q_t - p\cdot q}{2}
\edq
 From Theorem (3.5)
  of \cite{coro2} (and its proof)  we have the following propagation
estimates in the $L^2$-norm:
\medskip
\noindent  for every $N\in\N$ and every $T > 0$
there exists $C_{N, T}$  such that
\beq\label{propag}
\Vert U(t)\varphi_\alpha  -
 \exp\left(\frac{i \delta(\alpha, t)}{\hbar}\right){\cal T} (\alpha_t){\rm
Met}_\hbar(F(t))
 \Lambda_\hbar P_N(x, D_x, t, \hbar)\widetilde{\psi_0} \Vert \leq
C_{N, T}\hbar^N
 \edq
where $P_N(t, \hbar)$ is the $(\hbar,t)$-dependent differential operator
 defined by
\beqn
 P_N(x, D_x, t, \hbar) = \Id + \sum_{(k,j)\in I_N}\hbar^{k/2 - j}
p^w_{kj}(x, D, t) \nonumber \\
{ \rm with} \; I_N = \{(k,j)\in \N\times\N  ,\;1\leq j\leq 2N-1,\;k\geq 3j,\;
 1 \leq k-2j < 2N \}
\edqn
where the differential operators  $p_{kj}(x, D_x, t)$ are products of $j$
Weyl quantization
 of homogeous polynomials of degree $k_s$ with $\sum_{1\leq s\leq j}k_s =
k$ (see \cite{coro2}
Theorem (3.5) and its proof). So that we get
\beq
  p^w_{kj}(x, D_x, t) \widetilde{\psi_0} = Q_{kj}(x)\widetilde{\psi_0}(x)
\edq
where   $Q_{kj}(x)$ is a polynomial (with coefficients depending on
$(\alpha, t)$)
 of degree $k$ having the same parity as $k$. This is clear
 from the following facts: homogeneous polynomials have a definite parity,
and Weyl quantization
behaves well with respect to symmetries: $Op^w(A)$ commutes to the parity
operator $
\Sigma f(x) = f(-x)$
if and only if $A$ is an even symbol and  anticommutes with $\Sigma$  if
and only if $A$ is an odd symbol)
and $\widetilde{\psi_0}(x)$ is an even function.
So we get
\beqn\label{m1}
 m(\alpha, t) = \sum_{(j,k)\in I_N; \vert \gamma\vert \leq 2N}
 c_{k,j,\gamma} \hbar^{\frac{k+\vert \gamma\vert}{2}-j}
\exp\left(\frac{i\delta(\alpha,t)}{\hbar}\right)\cdot\nonumber\\ \cdot
\left\langle{\cal T}(\alpha)\Lambda_\hbar Q_\gamma \widetilde{\psi_0},
{\cal T}(\alpha_t)\Lambda_\hbar Q_{k,j}{\rm
Met}(F(t))\widetilde{\psi_0}\right\rangle + O(\hbar^{N})
\edqn
where $Q_{k,j}$ respectively $ Q_\gamma $ are polynomials in the $x$ variable
 with the same parity as $k$ respectively $\vert \gamma\vert$. This remark
will
be useful in proving that we have only entire powers in $\hbar$ in
(\ref{GTF}),
even though half integer powers appear naturally in the  asymptotic
propagation of coherent states.
By an easy computation we have
\beqn
\left\langle{\cal T}(\alpha)\Lambda_\hbar Q_\gamma \widetilde{\psi_0},
{\cal T}(\alpha_t)\Lambda_\hbar Q_{k,j}{\rm
Met}(F(t))\widetilde{\psi_0}\right\rangle
 = \nonumber \\
 \exp\left(-i\frac{1}{2\hbar}\sigma(\alpha,\alpha_t)\right)
\left\langle{\cal T}_1\left(\frac{\alpha-\alpha_t}{\sqrt{\hbar}}\right)
 Q_\gamma\widetilde{\psi_0},
Q_{k,j}{\rm Met}(F(t))\widetilde{\psi_0}\right\rangle
\edqn
 where ${\cal T}_1(\cdot)$  is the Weyl translation operator with $\hbar =
1$.\\
We set
 \beqn
m_{k,j,\gamma}(\alpha, t)  & = & \left\langle{\cal
T}_1\left(\frac{\alpha-\alpha_t}{\sqrt{\hbar}}\right)
 Q_\gamma\widetilde{\psi_0},
Q_{k,j}{\rm Met}(F(t))\widetilde{\psi_0}\right\rangle \\
m_0(\alpha, t) & = &  \left\langle{\cal
T}_1\left(\frac{\alpha-\alpha_t}{\sqrt{\hbar}}\right)
\widetilde{\psi_0},
 {\rm Met}(F(t))\widetilde{\psi_0}\right\rangle
\edqn
We compute $m_0(\alpha, )$ first.
We shall use the fact that the metaplectic group transforms Gaussian
wave packets to Gaussian wave packets in a very explicit way.
If we denote by $A$, $B$, $C$, $D$ the four $n
\times n$ matrices of the block form of $F(t)$,
\beq
F(t) = \pmatrix{A &\quad B \cr
C &\quad D \cr}
\edq
 it is clear, since $F$ is symplectic,  that
$U = A + iB$  is invertible. So we can define
\beq
M = VU^{-1},\;\;{\rm  where}\;\; V = (C + iD).
\edq
 We have (\cite{fo}, Ch.4)
\beqn\label{m0}
m_0(\alpha, t)  =  ({\rm det}\, U)_c^{-1/2}\pi^{-n/2}\cdot\nonumber\\
\cdot \int_{\R^n} \exp\left\{\frac{i}{2}(M+i\Id)x\cdot x)
 -\frac{i}{\sqrt{\hbar}}(x-\frac{q-q_t}{2})\cdot(p-p_t+i(q-q_t))\right\}dx
\edqn

\begin{remark} In \ref{m0}, $(z(t))_c^{1/2}$ has the following meaning: if $t
\mapsto z(t)$ is a continuous mapping from ${I\hskip-1truemm R}$ into
$\C\setminus\{0\}$ such that $z(0) > 0$ then $(z(t))_c^{1/2}$ denotes the
square root
defined by continuity in $t$ starting from $\sqrt{z(0)} > 0$. Thus factor (det
$U)_c^{-1/2}$ in (38) records the winding of det $U(t)$ at $t$ varies. This
takes the
place of the ``Maslov line bundle'' in this construction.
\end{remark}

If we make the change of variables $x \mapsto (y - q_t)/\sqrt{\hbar}$ in
(32) and
hence in (38), then the formula for the regularized density of states in
(19) takes
the form

$$\rho_A(E) = \int_R dt \int_{R^{2n}} d\alpha \int_{R^n} a(t, \alpha, y, \hbar )
e^{{i \over h} \Phi_E(y, \alpha , t)} dy  . \eqno(39)$$

\noi The phase function $\Phi_E$ is given by

$$\Phi_E(t,y, \alpha ) =$$
$$S(\alpha , t) + q \cdot p + (y -q_t)\cdot p_t + {1 \over 2} (y - q_t)
\cdot M(t)(y
- q_t) + {i \over 2} |y - q|^2 - y \cdot p + Et  , \eqno(40)$$

\noi where $\cdot$ denotes the usual bilinear product in $\C^n$, and
$\alpha = (q,
p)$, $\alpha_t = \phi_t(\alpha )$ as before. Our plan is to prove Theorem 2.3 by
expanding (39) by the method of stationary phase. The necessary stationary phase
lemma for complex phase functions can easily be derived from Theorem 7.7.5
in [22,
Vol. 1]. There is also an extended discussion of complex phase functions
depending
on parameters in [22] leading to Theorem 7.7.12, but the form of the stationary
manifold here permits us to use the following

\begin{theorem}[stationary phase expansion]  Let ${\cal O} \subset
{I\hskip-1truemm R}^d$ be an
open set, and let $a, f \in C^{\infty}({\cal O})$ with $\Im f \geq 0$ in
${\cal O}$
and supp $a \subset {\cal O}$. We define
$$
M = \{x \in {\cal O}, \Im f(x) = 0, f'(x) = 0 \}  ,
$$
 and assume that $M$ is a smooth, compact and connected submanifold of
${I\hskip-1truemm R}^d$ of dimension $k$ such that for all $x \in M$ the
Hessian, $f''(x)$, of $f$ is
nondegenerate on the normal space $N_x$ to $M$ at $x$.

 Under the conditions above, the integral $J(\omega ) = \int_{R^d} e^{i \omega
f(x)} a(x)dx$ has the following asymptotic expansion as $\omega \to + \infty$,
modulo $O(\omega^{- \infty})$,
$$
J(\omega ) \equiv \left ( {2 \pi \over \omega} \right )^{{d-k \over 2}} \sum_{j
\geq 0} c_j\omega^{-j}  .
$$
 The coefficient $c_0$ is given by
$$
c_0 = e^{i\omega f(m_0)} \int_M \left [ \det \left ( {f''(m)|N_m \over i}
\right )
\right ]_*^{-1/2} a(m) dV_M(m)  ,
$$
 where $dV_M(m)$ is the canonical Euclidean volume in $M$, $m_0 \in M$ is
arbitrary, and $[\det P]^{-1/2}_*$ denotes the product of the reciprocals
of square
roots of the eigenvalues of $P$ chosen with positive real parts. Note that,
since
$\Im f \geq 0$, the eigenvalues of ${f''(m)|N_m \over i}$ lie in the closed
right
half plane.
\end{theorem}

\noi {\bf Sketch of proof} : Using a partition of unity, we can assume that
${\cal
O}$ is small enough that we have normal, geodesic coordinates in a
neighborhood of
$M$. So we have a diffeomorphism

$$\chi : {\cal U} \to {\cal O}  ,$$
\noi where ${\cal U}$ is an open neighborhood of (0, 0) in
${I\hskip-1truemm R}^k \times
{I\hskip-1truemm R}^{d-k}$, such that

$$\chi (x', x'') \in M \Longleftrightarrow x'' = 0$$

\noi and if $m = \chi (x', 0) \in M$ we have

$$\chi '(x', 0)(R_x^k) = T_mM$$

$$\chi ' (x', 0)(R_{x''}^{d-k}) = N_m M , \hbox{normal space at $m \in M$)}
 . $$

\noi So the change of variables $x = \chi (x', x'')$ gives the integral

$$J(\omega ) = \int_{\hbox{{$I\hskip-1truemm R^d$}}} e^{i\omega f(\chi
(x', x''))} a(x', x'')|\det \chi '(x', x'')|dx'dx''  . \eqno(43)$$

\noi The phase

$$\tilde{f}(x', x'') \ := \ f(\chi(x', x''))$$

\noi clearly satisfies

$$\{\tilde{f}'_{x''}(x', x'') = 0, \Im \tilde{f}(x',x'') = 0 \} \Longleftrightarrow
x'' = 0  . \eqno(44)$$

\noi Hence, we can apply the stationary phase Theorem 7.7.5 of [22], (Vol.
1), in
the variable $x''$, to the integral (43), where $x'$ is a parameter (the
assumptions
of [22] are satisfied, uniformly for $x'$ close to 0). We remark that all the
coefficients $c_j$ of the expansion can be computed using the above local
coordinates and Theorem 7.7.5.

\section{The stationary Phase Computation}

In this section we compute the stationary phase expansion of (39) with
phase $\Phi_E$ given by (40).
Note that $a(t, \alpha , y, \hbar)$ is actually, according to (32), a
polynomial in
$\hbar^{1/2}$ and $\hbar^{-1/2}$. Hence the stationary phase theorem (with
$\hbar$
independent symbol $a$) applies to each coefficient of this polynomial. \\

The first order derivatives of $\Phi_E(t, y, \alpha )$ (up to $O((y -
q)^2, (\alpha - \alpha_t)^2)$ terms) are given by

\begin{eqnarray*}
&&\partial_t \ \Phi_E = E - H (\alpha ) + (y - q_t) \cdot \dot{p}_t -
\dot{q}_t \cdot
M(y - q_t) \\
&&\partial_y \ \Phi_E = p_t - p + i(y - q) + M(y - q_t) \\
&&\partial_q \ \Phi_e = i(q - q_t) - \ ^tA(p - p_t) +  (\ ^tC - \ ^tAM - i
I) (y -
q_t) \\
&&\partial_p \ \Phi_E = q - q_t + (\ ^tD - \ ^tBM - I) (y - q_t)  .
\end{eqnarray*}

\noindent Furthermore, since $F$ is symplectic, one has

$$2 {\rm Im} \ \Phi_E = |y - q|^2 + |(A + iB)^{-1}(y - q_t)|^2  .$$

\noindent This implies that $\Phi_E(y, \alpha , t)$ is critical on the set~:

$${\cal C}_E = \{(y, \alpha ,t) \in {I\hskip-1truemm R}_y^n \times
{I\hskip-1truemm
R}_{\alpha}^{2n} \times {I\hskip-1truemm R}_t \ : \ y = q_t, \ \alpha_t =
\alpha, \
H(\alpha) = E \}  .$$

\noi Thus each component ${\cal M}_\gamma$ of ${\cal C}_E$ has the form

$${\cal M}_\gamma = \left \{ (y, \alpha , t) = (q, \alpha , T(\alpha ))  :
\ \alpha = (p, q)
\in \gamma, \alpha_{T(\alpha )} = \alpha , H(\alpha ) = E \right \}  .
\eqno(45)$$
\noi We will assume that each $\gamma$ is a smooth compact manifold. One
sees immediately
that the manifolds $\gamma$ are unions of peridic classical trajectories of
energy $E$.
We will also assume a ``clean intersection" hypothesis which we will state
shortly.
Thus we have assumed that
$${\cal C}_E=\{ 0\}\times \Sigma \cup \{ {\cal M}_{\gamma_1},\dots,{\cal
M}_{\gamma_N}\},
\eqno{(46)}$$
where each ${\cal M}_{\gamma_k}$ has the form (45) with $\gamma_k$ in the
fixed point
set of the mapping $\alpha \mapsto \alpha_{T_k}$.

The first thing to check, in order to apply the stationary phase theorem is that
the support of $\alpha$ in (39) can be taken as compact, up to an error
$O(\hbar^{\infty})$. We do this in the following way: let us recall some
properties of $\hbar$-pseudodifferential calculus proved in
\cite{hero,disj}. The
function $m(z) = <H(z)>$ is a weight function. In \cite{disj} it is proved that
$\chi (\hat{H}) =  \hat{H}_{\chi}$ where $H_{\chi} \in S(m^{-k})$, for
every $k$.
More precisely, we have in the $\hbar$ asymptotic sense in $S(m^{-k})$,
$$H_{\chi} = \sum_{j \geq 0} H_{\chi j}\hbar^j$$

\noi and support $[H_{\chi , j}]$ is in a fixed compact set for every $j$
(see (H.5) and
\cite{hero} for the computations of $H_{\chi ,j})$. Let us recall that the
symbol space $S(m)$
 is equipped with the family of semi-norms
$$
\mathrel{\mathop {\rm sup}_{z \in \hbox{{$I\hskip-1truemm R^{2n}$}}}}
m^{-1}(z)|\frac{\partial^\gamma}{\partial z^\gamma}u(z)|
$$

\noi Now we can prove the following lemma

\begin{lemma} There is a compact set $K$ in $I\hskip-1truemm R^{2n}$ such
that for

$$m(\alpha ,t) = <\hat{A}_{\chi} \varphi_{\alpha}, U(t) \varphi_{\alpha}>$$

\noi we have

$$\int_{\hbox{{$I\hskip-1truemm R^{2n}$}}/K} |m(\alpha , t)|d\alpha =
O(\hbar^{+ \infty})  ,$$

\noi uniformly in every bounded interval in $t$.
\end{lemma}

\noi {\bf Proof:} Let $\tilde{\chi} \in C_0^{\infty}(]E - \delta E, E +
\delta E[)$ such
that $\tilde{\chi}\chi = \chi$. Using (H.4) and the composition rule for
$\hbar$-pseudodifferential operators  we can see that
$\hat{A}_{\chi}(\hat{H})$ is
bounded on $L^2({I\hskip-1truemm R}^n)$. So there exists a $C > 0$ such that

$$|m(\alpha , t)| \leq C \parallel \tilde{\chi}
(\hat{H})\varphi_{\alpha}\parallel^2  .$$

\noi But we can write

$$\parallel \tilde{\chi}(\hat{H}) \varphi_{\alpha}\parallel^2 =
<\tilde{\chi}(\hat{H})^2 \varphi_{\alpha}, \varphi_{\alpha}>  .$$

\noi Let us introduce the Wigner function, $w_{\alpha}$, for
$\varphi_{\alpha}$ (i.e. the Weyl symbol of the orthogonal projection on
$\varphi_{\alpha}$). We have

$$<\tilde{\chi}(\hat{H})^2 \varphi_{\alpha},\varphi_{\alpha}> = (\pi
\hbar)^{-n} \int H_{\chi^2}(z) w_{\alpha}(z) dz$$
\noi where

$$w_{\alpha} (z) = (\pi \hbar )^{-n} \ e^{-{|z - \alpha|^2 \over \hbar}}
 .$$

\noi Using remainder estimates from \cite{hero} we have, for every $N$
large enough,

$$\hat{H}_{\chi^2} = \sum_{0 \leq j \leq N} H_{\chi^2,h} \hbar^j +
\hbar^{N+1} R_N(\hbar )$$

\noi where the following estimate in Hilbert-Schmidt norm holds

$$\mathrel{\mathop {\rm sup}_{0 < \hbar \leq 1}} \parallel R_N(\hbar
)\parallel_{HS} < + \infty  .$$

\noi Now there is an $R > 0$ such that for every $j$, we have
Supp$[H_{\chi^2,j}]
\subseteq \{z, |z| < R\}$. So the proof of the lemma follows from

$$\parallel R_N(\hbar )\parallel_{HS}^2 = (2 \pi \hbar )^{-n} \int
\parallel R_N(\hbar) \varphi_{\alpha} \parallel^2 \ d\alpha,$$

\noi and from the elementary estimate, which holds for some $C$, $c>0$,

$$\int_{|z|\leq R,|\alpha|\geq R+1} e^{-{|\alpha - r|^2 \over \hbar}} dz d\alpha
\leq Ce^{-{c \over \hbar}}. \quad  \sq$$

\vskip 5 truemm
The next step is the computation of the Hessian of $\Phi_E$ on a ${\cal
M}_{\gamma k}$. After an easy but tedious computation, with the variables
ordered as
$(t, y, p, q)$, the Hessian $\Phi_E''$ is the following $(1 + 3n) \times (1
+ 3n)$
matrix

$$\Phi''_E = \left ( \begin{array}{cccc}
H_p \cdot (H_q + M H_p) &-H_q - H_p
M &-H_p(D-MB) &-H_p(C - MA) \\
-H_q - MH_p &M + iI &D - MB- I &C - MA - iI \\
-(^tD - \ {^t}B M)H_p &^tD - \ {^t}BM - I &^tB MB - \ {^t}DB &^tBMA - \
{^t}BC \\
-(^tC - \ {^t}AM)H_p &^tC - \ {^t}AM - iI &^tAMB - \ {^t}CB &^tAMA - \
{^t}CA + iI
\end{array} \right ) \eqno(47)$$

\noi where $H_p$ (resp. $H_q$) denotes the vector $\partial_pH|_{\alpha =
\alpha_t}$
(resp. $\partial_q H|_{\alpha = \alpha_t}$), $A$, $B$, $C$, $D$, are the $n
\times
n$ matrices given by (36), $^tA$ the transpose of $A$, and $M$ is defined
by (37).
(Recall $I$ is the identity matrix). \par

We are going to perform elementary row and column operations on (47)
to compute the
nullspace of $\Phi''_E$, and the determinant of $\Phi''_E$
restricted to the normal space to the critical
manifold. To begin with we have $H_1 = \ {^tR}_0 \Phi''_E R_0$ where

$$R_0 = \left ( \begin{array}{llll} 1 &0 &0 &0 \\ H_p &I &0 &0 \\ 0 &0 &I
&0 \\ 0
&0 &0 &I \end{array} \right )$$

\noi and $H_1$ is given by

$$H_1 = \left ( \begin{array}{cccc}
H_p \cdot (-H_q + iH_p) &- H_q + iH_p &-H_p &-iH_p \\
-H_q + i H_p &M + iI &D - MB - I &C-MA - iI \\
-H_p &^tD - \ {^t}BM - I &^tBMB - \ {^t}DB &^tBMA - \ {^t}BC \\
-iH_p &^tC - \ {^t}AM - i I &^tAMB - \ {^t}CB &^tAMA - \ {^t}CA + iI
\end{array}\right )$$

\noi Multiplying $H_1$ on the right by

$$R_3 = \left ( \begin{array}{llll} 1 &0 &0 &0 \\ 0 &I &B &A \\ 0 &0 &I &0
\\ 0 &0 &0
&I \end{array} \right )$$

\noi changes it to

$$H_2 = \left ( \begin{array}{cccc} H_p(-H_q + iH_p) &-H_q + iH_p
&-H_p + (-H_q + i H_p)B &(-H_q + iH_p)A - iH_p\\
-H_q + i H_p &M + iI &D-I + iB &C + i(A - I) \\
- H_p &^tD - \ {^t}BM - I &-B &I - A \\
-i H_p &^tC - \ {^t}AM - iI &-iB &-i(A - I)
\end{array}\right )$$

The key simplification comes from (37) which gives $M = (C + iD)(A
+iB)^{-1}$, and
hence, since $F$ is symplectic

$$^tD - \ {^t}BM = [^t D(A + iB) - \ {^t}B (C + iD)] (A + iB)^{-1} = (A +
iB)^{-1}$$
$$^tC - \ {^t}AM = [^t C(A + iB) - \ {^t}A(C + iD)] (A + iB)^{-1} = - i(A +
iB)^{-1}.$$

\noi Thus, subtracting the appropriate multiples of the third row in $H_2$ from the
other rows we get

$$H_3 = \left ( \begin{array}{cccc}
0 &(-H_q + iH_p)(A +iB)^{-1} &-H_p &-H_q \\ -H_q &(C + iD + iI) (A +
iB)^{-1} &D - I &C \\
- H_p &(I - A - iB) (A + iB)^{-1} &-B &I - A \\
0 &-2i(A + iB)^{-1} &0 &0 \end{array} \right ) . $$

\noi Finally using the fourth row to remove the three upper entries in the
second
column, multiplying the third row by $- 1$, interchanging the second and fourth
rows, and the third and fourth columns, we arrive at the simple form

$$H_4 = \left ( \begin{array}{cccc} 0 &0 &-H_q &-H_p \\ 0 &-2i(A +
iB)^{-1} &0 &0 \\ H_p &0 &A - I &B \\ -H_q &0 &C &D - I\end{array}\right )
.\eqno(48)$$

\noi and $H_4 = R_1 \Phi''_E R_2$ where $R_1$ and $R_2$ can be computed by
repeating
the elementary row and column operations that we have performed on the identity
matrix, and in particular det $R_1 = 1$ and det $R_2 = (-1)^n$. \par

In order to apply the stationary phase theorem the null space of $\Phi''_E$
must be
the tangent space to the critical set ${\cal C}_E$. However, one can read
off the null
space of $H_4$ from (48)

$$\hbox{Null} \ H_4 = R_2^{-1} \hbox{Null} \ \Phi''_E =$$
$$\left \{ (\tau , 0, v, w) : (F - I) {v \choose w}
 + \tau {H_p \choose - H_q}  = 0 \
\hbox{and} \  H_q \cdot v + H_p \cdot w = 0 \right \}  . \eqno(49)$$
This leads us to impose the following
``clean flow condition''
\vskip5mm
\noindent{\bf Hypothesis C:} Assume that
${\cal D}_E := \{ (\alpha , t) \in \Sigma_E \times {I\hskip-1truemm
R}\ /\phi_t(\alpha ) = \alpha \}$ is a submanifold of ${I\hskip-1truemm
R}^{1+2n}$.
Then we say the ${\cal D}_E$ satisfies the clean flow condition, if for
any $(\alpha , t) \in {\cal D}_E$, the tangent space to ${\cal D}_E$
is given by

$$T_{\alpha , t} \ {\cal D}_E = \left \{ (v,w, \tau ) \in {I\hskip-1truemm
R}^{1+2n}
: (F - I) {v \choose w} + \tau {H_p \choose - H_q} = 0 \ \hbox{and} \
H_q\cdot v +
H_p \cdot w =0 \right \} \ .\eqno(50)$$
\vskip5mm
 Since
${\cal C}_E = \left \{ (y , \alpha , t) : (\alpha , t) \in {\cal D}_E \
\hbox{and}
\ y = q \right \}$,
the tangent space $T_{y , \alpha , t} \ {\cal C}_E$ equals

$$\{(\tau , v, w, v) : (F - I) {v \choose w} + \tau {H_p \choose -H_q} = 0
\ \hbox{and} \ H_q \cdot v + H_p \cdot w = 0 \}, $$
and, assuming Hypothesis C, this does equal the null space of $\Phi''_E$, since
$$R_2 = \left ( \begin{array}{c} \tau \\ 0 \\ v \\ w \end{array} \right ) =
\left ( \begin{array}{c} \tau \\ Av + Bw + \tau H_p \\ w \\ v \end{array}
\right ) =
\left ( \begin{array}{c} \tau \\ v \\ w \\ v \end{array} \right )$$
for $(\tau, v, w )$ as in
(49).
Therefore, if $P$ denotes the orthogonal projection on the null space of
$\Phi''_E$, then $\det (\Phi''_E + P)$ will be the determinant of the Hessian of
the phase restricted to the normal space, and setting

$$\tilde{P} = R_1 P R_2 \eqno(51)$$

\noi we have $\det (H_4 + \widetilde{P}) = -(-1)^n \det (\Phi''_E + P)$.
Hence the computations of our paper provide a proof for the existence of a
Gutzwiller trace formula under Hypothesis C. However, as stated earlier, we
will only carry out the computations for the case that $\gamma$ consists
of a single trajectory here.
In this case Hypothesis C reduces to the assumption (H.3) of isolated
nondegenerate periodic orbits, and we may complete the computation in
the following way.

To compute det$(H_4 + \tilde{P})$ we will use a special basis ${\cal B}$.
We denote
by $E_{\lambda}$ the (algebraic) eigenspace of $F$ belonging to the eigenvalue
$\lambda$. Then under assumption (H.3)

$${\rm dim} \mathrel{\mathop \oplus_{\lambda \not= 1}} E_{\lambda} = 2n - 2$$
$${\rm dim} \ E_1 = 2$$

\noi and $\sigma (E_{\lambda}, E_1) = 0$ for $\lambda \not= 1$ where $\sigma$ is
the symplectic form, as in (10).
Let $(z_1, z_2)$ be a basis for $E_1$
with

$$z_1 = (2H_p^2 + H_q^2)^{-1/2} (H_p, - H_q),  $$
and $(F-I)z_2=\beta z_1$. Let $m_1, \cdots m_{2n-2}$ be a
(real) basis for the span of $\displaystyle{\mathrel{\mathop
\oplus_{\lambda\not=1}}}
E_{\lambda}$, and let $e_0, \cdots e_n$ be the Euclidean basis for
${I\hskip-1truemm R}^{n+1}$.
Then we take  ${\cal B}$ to be the basis

$$\{ (e_0, 0) \cdots (e_n, 0) \} \cup \{(0, m_1) \cdots , (0, m_{2n-2})\} \cup
\{ (0, z_1), (0, z_2)\} .$$

\noi Since the vector $\tilde{P} {0 \choose z_1}$ spans the range of
$\tilde{P}$ and
$H_4 {0 \choose z_1} = 0$, we can use column operations to remove the
contribution of $\tilde{P}$ from all columns of the matrix $H_4 + \tilde{P}$
with respect to ${\cal B}$, except the one corresponding to $z_1$. Then
we can use column operations to
remove all entries in the $z_1$- and $z_2$-columns corresponding to
the basis vectors $(e_1, 0) \cdots (e_n, 0)$, and $(0,m_1) \cdots (0,
m_{2n-2})$. Note that this does not change the entries in the first row of the
matrix, since $\sigma(z_1,m_j)=0,\ j=1,\dots 2n-2$.
After these simplifications which do not change the determinant, the
matrix of $H_4 + \tilde{P}$ with respect to ${\cal B}$ becomes:

$$\left ( \begin{array}{cccc} 0 &0 &0 &b \cr
0 &-2i(A + iB)^{-1} &0 &0 \\
0 &0 &P_{\gamma} - I &0 \\
a &0 &0 &\Omega \cr \end{array} \right )  \eqno(52)$$

\noi The vector $a$ is just $((2H_p^2 + H_q^2)^{1/2}, 0, \cdots 0)$ and

$$b = \left ( x , -(2H_p^2 + H_q^2)^{1/2} \sigma
(z_1, z_2) \right ) \quad .$$

\noi Therefore the determinant of $H_4 + \tilde{P}$ equals

$$ (-i)^n \left [ \det {A + iB \over 2} \right ]^{-1} \det (P_{\gamma}
- I) \det \tilde{\Omega},\eqno(53)$$

\noi where

$$\tilde{\Omega} = \left ( \begin{array}{ll} 0 &b \\ a &\Omega \end{array}
\right ) =
\left ( \begin{array}{ccc} 0 &x &-(2H_p^2 + H_q^2 ) \sigma (z_1,
z_2) \\ (2H_p^2 + H_q^2)^{1/2} &x &x \\
0 &c &0 \end{array}\right ).  \eqno(54)$$

\noi Here $x$ is used for entries that do not enter the calculation, and $c$ is the
component of $\tilde{P}(0, z_1)$ along the basis vector $z_2$. \par

To compute $c$ and finish the computation of the determinant, we
first compute $\tilde P(0,z_1)$. Writing $z_1=(v,w)$, we have
$$\tilde{P} \left ( \begin{array}{c} 0 \\ 0\\ v \\ w \end{array} \right ) =
{1 \over 2}
\left ( \begin{array}{l} x \\ x \\ - w + Bv - Aw \\ 3v - Cw + Dv
\end{array} \right
).  \eqno(55)$$

\noi We let $\tilde{P}_1z_1$ denote the last $2n$ components of
$\tilde{P}(0,z_1)$. Since $^tFJz_1=Jz_1$, the normalization in the
definition of $z_1$ gives, $\sigma(z_1,\tilde P_1z_1)=1$.
Therefore, if $\tilde{P}_1z_1 = c z_2 + dz_1$ we clearly have $c = \sigma
(z_1, z_2)^{-1}$. Thus (54) yields

$$\det \tilde{\Omega} = - (2H_p^2 + H_q^2) \eqno(56)$$

\noi and, combining this with (53) and (56), we have

$$\det \Phi''_E |_{N({\cal M}_{\gamma})} = (-1)^{n-1} (-i)^n \det \left
( {U \over 2} \right )^{-1} |(0, H_p , - H_q, H_p)|^2 \cdot \det
(P_{\gamma} - I) . \eqno(57)$$

Using (42) and (57), we conclude

\begin{eqnarray*}
&&d_0^{\gamma} = \\
&&\hat{g} (T_{\gamma}) e^{iS_{\gamma}/\hbar} \int_0^{T^*_{\gamma}} \left [
{(-1)^{1
- n} |(0,\dot{q}_s, \dot{p}_s, \dot{q}_s)|^2 \det ({\cal P}_{\gamma} - I)
\over \det
\left ( {U \over 2} \right )} \right ]^{-1/2}_* \left ( \det {U \over 2} \right
)_c^{-1/2} A(\alpha_s) dV(s) . \end{eqnarray*}

\noi Using $|(0, \dot{p}_s, \dot{q}_s, \dot{p}_s)|^{-1}dV(s) = ds$ we get the
result for $d_0^{\gamma}$ in (13). Since $\det ({\cal P}_{\gamma} - I) =
(-1)^{\sigma '} |\det ({\cal P}_{\gamma} - I)|$, where $\sigma '$ is the
number of
real eigenvalues of ${\cal P}_{\gamma}$ which are greater than 1, we see that

$$\left [ {(-1)^{1-n} \det ({\cal P}_{\gamma} - I) \over \det \left ( {U \over
2} \right )} \right ]^{-1/2}_* \left ( \det {U \over 2} \right )_c^{-1/2} = \pm
i^{n-1 + \sigma '} | \det ({\cal P}_{\gamma} - I)|^{-1/2}  . \eqno(58)$$

\noi Note that the role of the Maslov index in (13) is to determine the sign in
(58) and $\sigma_{\gamma}$ in (13) is either $n - 1 + \sigma '$ or $n + 1 +
\sigma '$.

The other coefficients, $d_j^{\gamma}$ are spectral invariants which have been
studied by Guillemin and Zelditch. In principle we can compute them using this
explicit approach. This completes the proof of Theorem 2.3.\\

\end{document}